\documentclass[prb,11pt]{revtex4-1}
\pdfoutput=1

\usepackage{amsmath}    
\usepackage{graphicx}   
\usepackage{verbatim}   
\usepackage{color}      
\usepackage{subfigure}  
\usepackage{hyperref}   
\usepackage[retainorgcmds]{IEEEtrantools}  
\usepackage{tikz}      
\usepackage{pgfplots}
\usetikzlibrary{shapes.symbols}
\usepgflibrary{shapes.symbols}
\usepgflibrary{shapes.geometric}
\usetikzlibrary{calc}
\usepackage[pagewise]{lineno}
\usepackage{tikz-3dplot}

\newcommand{\ud}{\, \mathrm{d}}

\newcommand{\h}{\mathbf{H}}
\newcommand{\dmdt}{\frac{\ud\m}{\ud t}}
\newcommand{\m}{\mathbf{m}}

\newcommand{\heff}{\h_{\mathrm{eff}}}
\newcommand{\unit}[1]{\ \mathrm{#1}}

\begin{document}

\title{Pinned domain wall oscillator as tunable direct current \mbox{spin wave} emitter}
\author{Michele Voto, Luis Lopez-Diaz, Eduardo Martinez}
\affiliation{Departamento de F\'isica Aplicada, Universidad de Salamanca, Plaza de la Merced s/n. 37008 Salamanca, Spain}

\begin{abstract}
Spin waves are perturbations in the relative orientation of magnetic moments in a continuous magnetic system, which have been proposed as a new kind of information carrier for spin-based low power applications. 
For this purpose, a major obstacle to overcome is the energy-efficient excitation of coherent short wavelength spin waves and alternatives to excitation via the Oersted field of an alternating current need to be explored. 
Here we show, by means of micromagnetic simulations, how, in a perpendicularly magnetized thin strip, a domain wall pinned at a geometrical constriction emits monochromatic spin waves when forced to rotate by the application of a low direct current flowing along the strip.
Spin waves propagate only in the direction of the electron's flow at the first odd harmonic of the domain wall rotation frequency for which propagation is allowed. 
Excitation is due to in-plane dipolar stray field of the rotating domain wall and that the resulting unidirectionality is a consequence of the domain wall displacement at the constriction. 
On the other hand, the application of an external field opposing domain wall depinning breaks the symmetry for spin wave propagation in the two domains, allowing emission in both directions but at different frequencies. 
The results presented define a new approach to produce tunable high frequency spin wave emitters of easy fabrication and low power consumption.
\end{abstract}

\flushbottom
\maketitle

\thispagestyle{empty}

A spin wave (SW) is a propagating perturbation in the magnetic texture in the form of a phase-coherent precession of the magnetic moments in a magnetic medium\cite{Stancil2009,Kruglyak2010,Karenowska2015}. The quanta of spin waves are called magnons and the field of science investigating the transmission and processing of information mediated by spin waves is termed magnonics. 
Magnonics offers a promising new route for computing technology, since it may overcome the limitations of complementary metal oxide semiconductor (CMOS) technology in terms of scalability and power consumption via a particle-less transmission of information\cite{Stamps2014,Chumak2014,Vogt2014,Khitun2010,Schneider2008,Klingler2015} and introducing new degrees of freedom encoded in spin waves' transport of angular momentum. 
Spin waves have short wavelengths at the technologically relevant$\unit{GHz}$ - low$\unit{THz}$ frequencies, allowing for integration with microwave electronics at the nanoscale\cite{Kruglyak2010}.
The classical technique used to inject spin waves is via the Oersted field induced around a wire placed on top of the spin waves conduct by an ac current flowing through it\cite{Schneider2008,Vogt2014,Chumak2012}. This approach allows control of frequency and wavelength of injected magnons, with the main drawback  that antenna width sets a lower bound for wavelength and limits the scalability of the device.
The conversion of electron-carried angular momentum into magnons\cite{Demidov2010,Madami2011,Demidov2012} and vice-versa\cite{Kajiwara2010,Chumak2012} allows for the exploitation of  spintronics phenomena for generation and detection of spin waves at the nanoscale and the embedding of magnonic circuitry in electronic-based devices. This novel field is called magnon-spintronics. 
Here, spin waves generation can be achieved by various localized excitations, such as electric field control of magnetostrictive properties of materials\cite{Rovillain2010,Weiler2011,Cherepov2014}, spin transfer torque\cite{Berger1986,Slonczewski1996a,Zhang2004} (STT) based spin waves excitation, generated by a spin-polarized current flowing through a nanocontact\cite{Demidov2010,Madami2011} or via the spin current originated by the flowing of charge current through an adjacent non-magnetic metal with large spin-orbit coupling\cite{Demidov2012,Liu2013,Hamadeh2014,Awad2016}. 
The use of an oscillating domain wall (DW) as a tunable spin wave emitter, excited by an alternate current, has been proposed by Van de Wiele and colleagues~\cite{VandeWiele2016}; in their work, a strong pinning for the DW is achieved via ferromagnetic-ferroelectric coupling and ac current is used to generate DW oscillations that excite propagation of SWs in adjacent domains at an angle of $45^\circ$ with respect to magnetization orientation. The use of a DW permits SW excitation at wavelengths much shorter than what can be achieved with common antennas and the change in ac frequency can, to some extent, regulate the emitted frequency. However, the realization of such device presents some limitations: high current densities are needed, the fabrication of hybrid ferromagnetic-ferroelectric structures and the non trivial propagation of SW in the $45^\circ$ magnetized domains.

On the other hand, it was shown both analytically and numerically\cite{Ono2008,Bisig2009,Martinez2011,Ghosh2014}, that a DW pinned at a constriction in a perpendicularly magnetized nano-wire could 
be led to self-sustained full in-plane rotation by the STT exerted on it by the application of a low in-plane dc current  while remaining pinned at a localized pinning site, thus creating a DW-based oscillator with frequencies in the GHz range, tunable via the applied current intensity. 
In this work we reconsider the DW oscillator set-up, schematically represented in Fig.~\ref{Fig:1}-a, and investigate, using micromagnetic simulations, the emission of SW generated by such localized magnetization precession in a nano-wire.
By selecting an adequate wire width and constriction geometry we can achieve a wide operating window in which we observe DW rotation at a current dependent frequency $f_{DW}$ that leads to unidirectional emission of SWs in the direction of electrons' flow at odd  harmonics of DW frequency. Since the SW frequency is a multiple of $f_{DW}$ we can tune the frequency of the emitted SWs with current as well. Moreover, via the application of an external field opposing the force exerted by the current, the device operating window is extended and, at the same time, the symmetric dispersion relation for SW in the two antiparallel domains is naturally split, which allows us to selectively propagate different harmonics in each domain. We identify the DW's in-plane stray field as the main responsible for SW excitation, whereas the unidirectionality is due to the asymmetric position of the DW below the geometrical constriction. This new concept of SW emitter has the attractive features of high coherence, tunable frequency up to tens of GHz and low power consumption (typical current of a few $\unit{\mu A}$), by simply exploiting the stray field induced by geometrical patterning.

\section*{Results}
\begin{figure}[b]
\includegraphics[width=\textwidth]{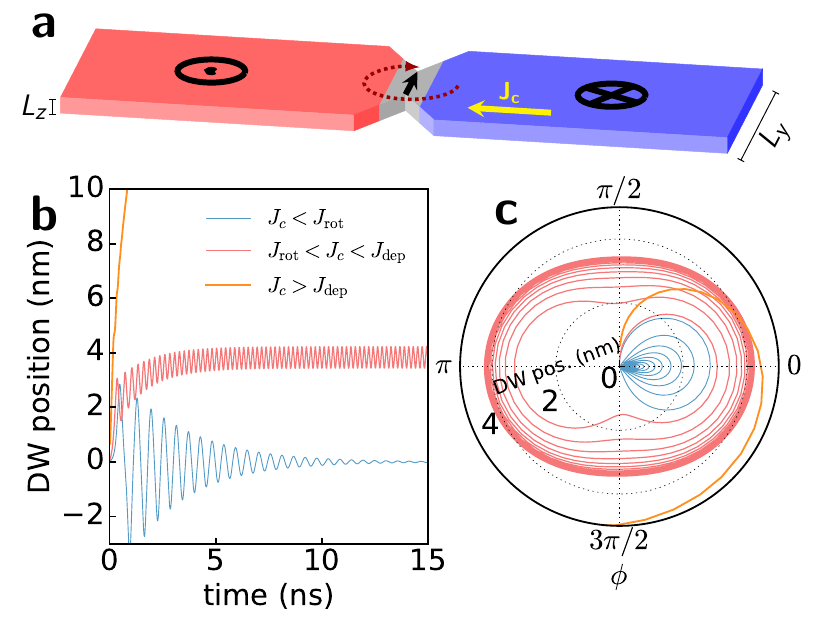}
\caption{\textbf{a} Schematic of the system under study: a DW situtated at a symmetric notch separates up (red) and down (blue) domains. Current flows from right to left so that the induced rotation of the DW is clockwise. In-plane direction of rotation is marked. \textbf{b} DW position as funciton of time in micromagnetic simulations with  current values below (dark blue line), inside (purple line) and above (orange line) the pinned rotation window. \textbf{c} Data from same simulations as in b plotted to represent DW position $q$ as the radial coordinate and DW angle $\phi$ as polar coordinate. }\label{Fig:1}
\end{figure}

We consider a DW trapped at a symmetric constriction in a narrow wire and a dc current flowing through it as shown in Fig.~\ref{Fig:1}-a. The constriction acts as a pinning site for the DW and, therefore, a minimum threshold current $J_\mathrm{dep}$ is needed to depin the DW and propagate it through the nanowire. For current densities below this threshold, the DW remains pinned at a position where the restoring pinning force balances the driving STT force that pushes the DW away from the notch. However, a zero net driving force does not imply balance on the in-plane torques acting on the DW. In particular, if a current density is above a certain value $J_\mathrm{rot}$, the in-plane component of STT overcomes the shape anisotropy field torque~\cite{Koyama2011}, leading to sustained full in-plane rotation of the spins inside the DW~\cite{Bisig2009,Martinez2011}.

Such situation requires the fulfilment of precise conditions that we synthetically present below using a typical one dimensional model~\cite{Schryer1974,Mougin2007a,Thiaville2004} (see Methods) that provides a good approximation to complex DW dynamics in narrow wires. Within this model the system is described using the position of the DW $q$ with respect to its equilibrium position centred at the constriction and the in-plane orientation $\phi$ of the spins in the DW as the only degrees of freedom. The pinning due to the constriction is modelled as a parabolic potential well\cite{Martinez2011} that gives rise to a spring-like restoring field $H_p(q)$. Within this framework, pinned rotation of the DW is attained for current density values $J_c$ in the range

\begin{equation}
J_\mathrm{rot}=\frac{e M_s}{P \mu_B}\frac{\gamma_0\Delta H_K}{2}< J_c < \frac{e M_s}{P \mu_B}\frac{\alpha \gamma_0\Delta H_p^{max}}{1+\alpha\beta}=J_\mathrm{dep},\label{window0}
\end{equation}

where all the quantities are defined in Methods, while the analytical derivation of such conditions can be found in section A of supplementary material.

Data from micromagnetic simulations realized for the three different cases $J_{c,1}<J_\mathrm{rot}<J_{c,2}<J_\mathrm{dep}<J_{c,3}$ are shown in Fig.~\ref{Fig:1}-b, where DW position is plotted as function of time, and in Fig.~\ref{Fig:1}-c, where the position and in-plane DW angle are shown in polar coordinates.
As can be observed, the DW reaches an equilibrium position for $J_c<J_\mathrm{rot}$ after a few nanoseconds, whereas for $J_c>J_\mathrm{dep}$ the DW rapidly depins from the notch. For $J_\mathrm{rot} < J_c < J_\mathrm{dep}$, however, the DW moves a few nm towards the right and slightly oscillates back and forth around this position while rotating in-plane.

In order to have a large operating window of the device we tune the wire width and notch shape to obtain a low threshold current for the DW pinned rotation $J_\mathrm{rot}$, and a high threshold current for DW depinning $J_\mathrm{dep}$. 
Since $|\mathbf{J}_c|$ value is not constant in space and increases at the constriction, throughout the paper we refer to its value as the nominal one away from the geometrical constriction. We select a wire width $L_y$ of 60 nm, a thickness  $L_z=1\unit{nm}$ and a notch depth of 20 nm, which gives us $J_\mathrm{rot}=10^{10}\unit{Am^{-2}}$ and $J_\mathrm{dep}=12.75\times 10^{10}\unit{A\,m^{-2}}$. The latter corresponding to a maximum current intensity of $7.6 \unit{\mu A}$ through the nanowire. The working window of such device has the desirable quality of lying in a low current density range, which allows us to avoid Joule heating effects and significant temperature gradients in proximity of the constriction~\cite{Moretti2016}.

\subsection*{Spin Wave Emission}
\begin{figure}[b]
\includegraphics[width=\textwidth]{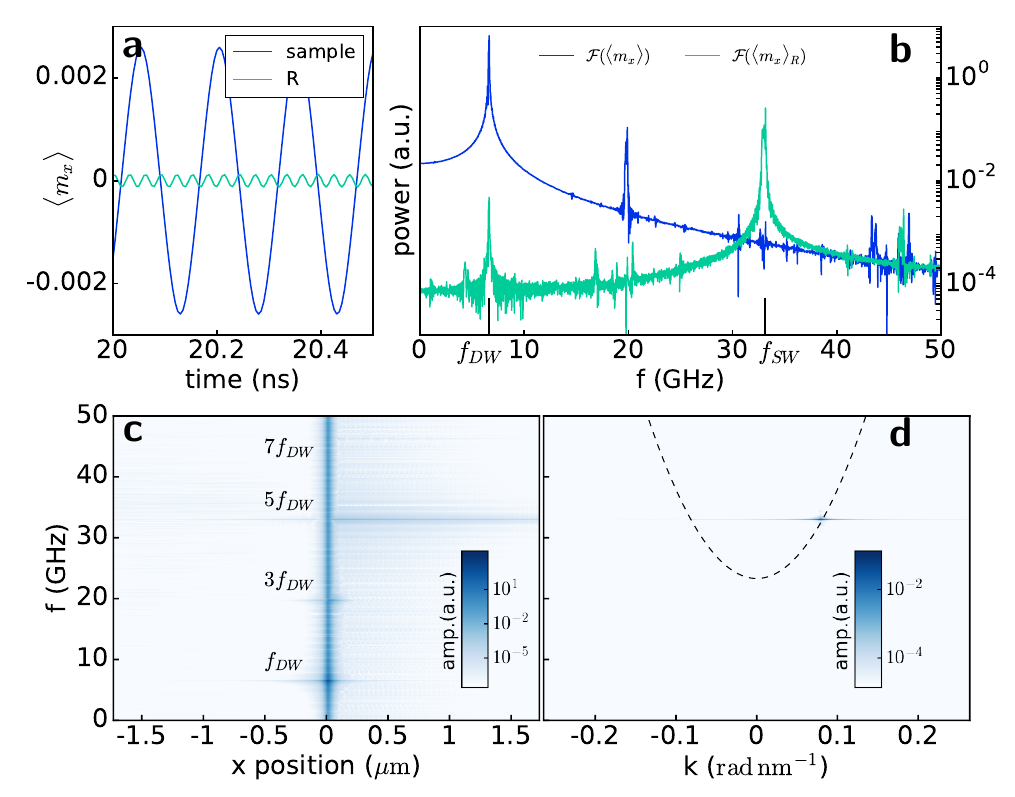}
\caption{\textbf{a} Evolution of the averaged $x-$ component of the normalized magnetization during a time window of $0.5\unit{ns}$. Dark blue line shows averaging over the whole sample, light blue line shows the averaging over the region $R$ away from the DW.  Different periodicity can be observed. \textbf{b} Fourier transfor of the time signals shown in \textbf{a}, dark blue line shows the main peak at the freqeuncy of rotation of the DW $f_{DW}$, light blue line has the principal peak at SW propagation frequency $f_{SW}=5f_{DW}$. \textbf{c} Frequency spectrum of $m_x(t)$ as function of $x-$ position along the line of micromangetic cells running at the center of the strip width. Peaks centred at the DW position with odd multiple frequency of $f_{DW}$ are marked.\textbf{d} $f-k$ diagram extracted from the central line in region $R$ shows a focused peak lying over the right principal branch of the analytical dispersion relation makerd as dashed line.}\label{Fig:3}
\end{figure}

Upon the application of a current $J_\mathrm{rot} < J_c < J_\mathrm{dep} $ through the wire, the domain wall is driven towards the right and, after a short time $\tau\leq10\unit{ns}$ that depends on the applied current, reaches a stationary position below the notch, where it slightly oscillates back and forth (see Fig.~\ref{Fig:1}-b) while its spins rotate clockwise in the strip plane as shown in Fig.~\ref{Fig:1}-c and in movie A of supplementary material. Due to the reduced lateral dimension of the wire, DW rotation is coherent and its spins rotate synchronously. Looking at the normalized $x$- component of magnetization $m_x=M_x/M_s$, as shown in Fig.~\ref{Fig:2}-a and in movie A in supplementary material for $J_c=6.5\times 10^{10} \unit{A\,m^{-2}}$, 
we observe the presence of the characteristic pattern of SW propagating to the right side of the strip whereas a much weaker propagation is observed on the left side. 
In Fig.~\ref{Fig:3}-a we monitor the value of the $x$- component of magnetization averaged over the whole strip, $\langle m_x \rangle$ (dark blue line). We observe its value oscillate around zero at the frequency $f_{DW}= 6.6 \unit{GHz}$. If we look at the average over a region $R$ $1.35\unit{\mu m}$ long, situated 400 nm away from the DW (light green line) we observe a smaller oscillation with higher frequency. By taking the Fourier transform of these two signals (Fig.~\ref{Fig:3}-b) we observe a peak  at $f_{DW}=6.6\unit{GHz}$, while the main signal from the region $R$ represents the SW frequency $f_{SW}=33\unit{GHz}$. Secondary peak in the global signal in correspondence of $3f_{DW}$ represents an odd higher harmonic.
In order to have more insight into the magnetization dynamics we look at the frequency signal distribution over space (Fig.~\ref{Fig:3}-c) by taking the Fourier transform of $m_x(t,\mathbf{r})$ at every cell situated along the $x$- central axis of the strip. Large amplitude can be observed at the centre of the strip where the DW rotates remaining pinned below the notch, with the largest amplitude at the frequency at which the DW rotates fully in-plane. Additional peaks at odd multiples of $f_{DW}$ can be seen with a propagating branch in correspondence with the fifth harmonic, indicating a definite propagation of SW towards the right. Taking the Fourier transform in space and time of $m_x(t,\mathbf{r})$  in the same central row of cells restricted to region $R$ we obtain the $f-k$ diagram showing a single focused spot in correspondence of $f_{SW}$ (Fig.~\ref{Fig:3}-d).The analytical dispersion relation for exchange spin waves in our sample is also shown in the figure
\begin{equation}
\omega(\mathbf{k})  =  \omega_0+\omega_M\lambda_\mathrm{ex}^2\mathbf{k}^2,
\label{disRel}
\end{equation}
where $\omega_0 = \gamma_0\left(H_\mathrm{k,eff}+H_a\right)$, $H_\mathrm{k,eff}$ is the effective out of plane anisotropy,  $\omega_M  =  \gamma_0 M_s$ and $\lambda_\mathrm{ex}^2  =  \frac{2 A}{\mu_0 M_s^2}$. 
Indeed $f_{SW}=5f_{DW}$ is the first odd harmonic that is allowed to propagate in the system, being above the threshold frequency $f_0=\omega_0/2\pi=23.3\unit{GHz}$. This emission of SW has the remarkable property of being unidirectional, coherent and directly dependent on the applied current density as we will discuss below.

Varying the applied current intensity between $J_\mathrm{rot}$ and $J_\mathrm{dep}$ leads to different DW  rotation frequencies extracted from $\mathcal{F}(\langle m_x \rangle)$ as shown in Fig.~\ref{Fig:4}-a and b. It is predicted by the analytical model\cite{Ono2008} that a linear relationship exists between $f_{DW}$ and applied current. However, DW rotation position varies with applied current, thus changing local current density at the DW position, so that DW rotation frequency is not linear with nominal applied current. Linear dependence between applied current and $f_{DW}$ is recovered if we consider the actual current density flowing at the DW position (see supplementary material). Also the amplitude of the signal increases with current density as denoted by the size of the hexagons in Fig.~\ref{Fig:4}-b. 
This is due to the fact that $ \langle m_x\rangle $ oscillation amplitude increases with $J_c$ since the DW moves further away from the center of the notch and consequently its length increases.

If we now look at the frequency spectrum  away from the notch, we observe a different distribution of amplitude peaks with $J_c$. In Fig.~\ref{Fig:4}-c peaks in the frequency signal sampled in region $R$ are plotted against the applied current density with different color and size to mark their amplitude. Dashed blue lines denote the frequency-current density curve of the rotating DW and its odd higher harmonics. As can be observed, all peaks lie on odd harmonics of $f_{DW}$ and their amplitude is maximum when SW can actually propagate, i.e above the threshold frequency $f_0$. The frequency gap region where propagation is forbidden is shaded in blue.
The highest emission intensity is achieved for current densities between 4.5 and $7\times 10^{10}\unit{A\,m^{-2}}$ on the fifth harmonic which is the first branch largely above the propagation threshold $f_0$. Emission is highly coherent with linewidths below 150 MHz.

\subsection*{Application of external field}
In order to extend the operating window of the device, we apply an external field opposing the driving torque due to STT. In our situation this means applying an external field $H_a$ pointing into the plane along $-\hat{\mathbf{z}}$ direction. We can estimate such effect by means of one dimensional model: the external field together with the pinning effective field have to balance the STT so that we have a linear dependency of depinning current from external field.
\begin{equation}
J_\mathrm{dep}(H_a)=\frac{e M_s}{ P \mu_B}\frac{\alpha\gamma_0\Delta}{1+\alpha\beta}\left(H_p(\overline{q})+H_a\right)=J_\mathrm{dep}^0+\frac{e M_s}{P \mu_B}\frac{\alpha\gamma_0\Delta}{1+\alpha\beta}H_a,\label{1Dthreshold}
\end{equation}
where $J_\mathrm{dep}^0$ is the threshold current for depinning without an applied field from equation~\eqref{window0}.

The increase in the depinning current in presence of an external field is shown in Fig. \ref{Fig:5}-a together with the analytical prediction~\eqref{1Dthreshold} (dashed line). Threshold current $J_\mathrm{dep}$ increases almost linearly with applied field for a wide range of fields, with $120\unit{mT}$ giving a $100\%$ increment of the depinning current at zero field. This way we can extend the current density window for  DW rotation and achieve $f_{DW}$ up to  $15.7 \unit{GHz}$ as shown in Fig.~\ref{Fig:5}-c. An interesting consequence of the application of an into-the-plane field is its antisymmetric contribution to the effective field in the two magnetic domains in which our strip is divided, which leads to the vertical displacement of the left and right propagating branches (Fig.~\ref{Fig:5}-b) depending on the relative orientation with the magnetization. 
This splitting of the dispersion relation on the left and right domains has the consequence of opening the possibility of strong SW propagation also in the left domain, in fact, for an applied field of $300 \unit{mT}$ $f_0=23.3 \unit{GHz}$ becomes  $f_0^R=31.4 \unit{GHz}$ and $f_0^L=14.6 \unit{GHz}$, which is already in the $f_{DW}$ range.
We perform simulations on a longer strip of $8.192 \unit{\mu m}$, with an external applied field $-H_a\hat{\mathbf{z}}$ with $\mu_0 H_a=300 \unit{mT}$ and we monitor  the magnetization in two regions $L$ and $R$, $2 \unit{\mu m}$ long situated $1.296 \unit{\mu m}$ away from the centre of the strip as shown in Fig.~\ref{Fig:2}-b. The DW rotation frequency is extracted as usual as $\mathcal{F}(\langle m_x\rangle)$ and plotted in Fig.~\ref{Fig:5}-c.
In Fig.~\ref{Fig:5}-d the peaks in frequency of $m_x(t)$ in the two sampled regions are plotted, where symbols' size represent their amplitude.  
In the left domain (triangles) we have SW propagation towards the left at the DW rotation frequency when this exceeds the propagation threshold frequency $f_0^L$ and no higher harmonic excitation is observed, whereas on the right side, the third harmonic is now accessible for SW propagation due to the increased DW rotation frequency and is the one at which SW propagate towards the right.
This result adds an important feature to this spin wave emitter, since 
spin wave propagation can be tuned in two different aspects: propagation frequency can be regulated by changing applied current, while bidirectional or unidirectional emission from the DW and additional frequency regulation can be selected via the application of an external field.

\section*{Discussion}
This novel scheme for tunable, short wavelength SW emission can open new paths in low power magnonic devices.
Unidirectional and asymmetric spin waves propagation is a peculiar characteristic of this system. The intrinsically asymmetric character of Dzyaloshinskii-Moriya interaction (DMI) has been exploited to obtain unidirectional propagation of SW along nanowires~\cite{Bracher2016} and focusing of SW in thin films~\cite{Kim2016}. In our system however the effect of DMI is negligible and the origin of such effect is purely geometric as will be shown below.
The precession of the domain wall's spins is the source of spin waves excitation, and 
SW propagation at odd higher harmonic of DW rotation frequency is the signature of a periodic and non-linear excitation\cite{Wang2015a,Xia2016}.
If the simple oscillation of the DW below the notch or its change in width when passing from N\'eel to Bloch were the main mechanism of excitation, we would observe emission at $2f_{DW}$ and its harmonics. However, the absence in the frequency spectrum of amplitude peaks at even multiples of $f_{DW}$ makes us discharge this hypothesis.

\begin{figure}[b]
\includegraphics[width=\textwidth]{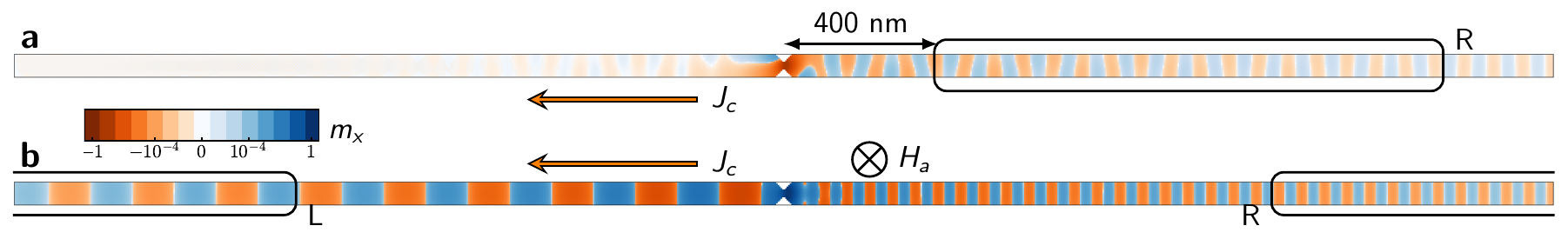}
\caption{\textbf{a} Snapshot of magnetization dynamics representing $M_x/M_s$ value when a current density of $6.5\times10^{10}\unit{A\,m^{-2}}$ is applied. The region $R$ on the right where $m_x$ is sampled, is enclosed by a rectangle. Unidirectional SW propagation towards the right can be observed. \textbf{b} Snapshot of the magnetization dynamics under the concurrent action  of an in-plane current of $24\times10^{10}\unit{A\,m^{-2}}$ and an external field of $300 \unit{mT}$ directed inside the plane to oppose DW depinning.  A wire twice as long as in \textbf{a} is considered. Emission of SW is observed both towards the right and the left at different frequencies and wavelengths, the sampled regions on the left and right are enclosed by black rectangles and they extend for $2\unit{\mu m}$.}\label{Fig:2}
\end{figure}

\begin{figure}[b]
\includegraphics[width=\textwidth]{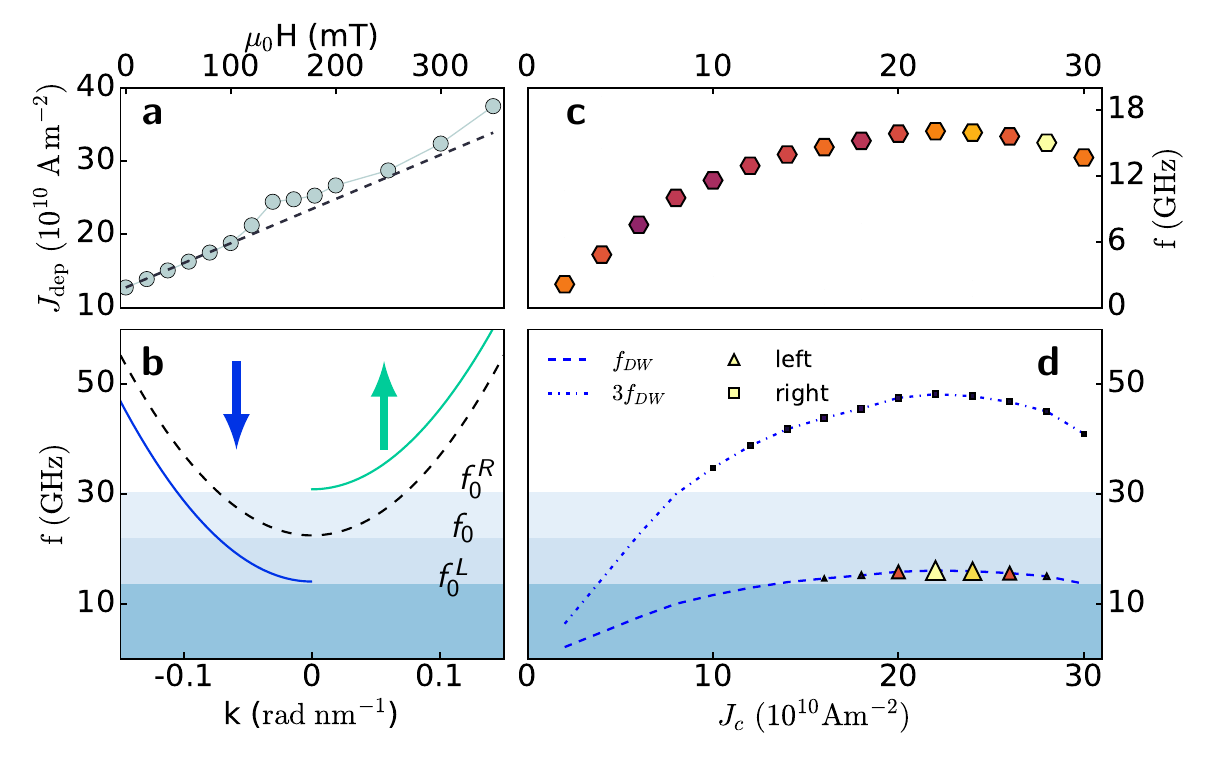}

\caption{\textbf{a} Depinning current $J_\mathrm{dep}$ as a function of a counter-acting applied field applied along $-\hat{\mathbf{z}}$ direction. The dashed line represents the theoretical prediction made using the one dimensional model~\eqref{1Dthreshold}. \textbf{b} Splitting of the dispersion relation branches in the positive and negative direction with respect to the DW due to the applicaiton of an external field as prescribed by equation~\eqref{disRel}. \textbf{c} DW rotation frequency as function of applied current when an external field of $300 \unit{mT}$ is applied. \textbf{d} SW propagation observed in the regions distant from the DW on the left (triangles) and on the right (squares) as indicated in Fig.~\ref{Fig:2}-b. Size of the symbols express the SW amplitude. SW  propagate to the left at $f_{DW}$ while to the right at $3f_{DW}$.}\label{Fig:5}
\end{figure}

\begin{figure}[b]
\includegraphics[width=\textwidth]{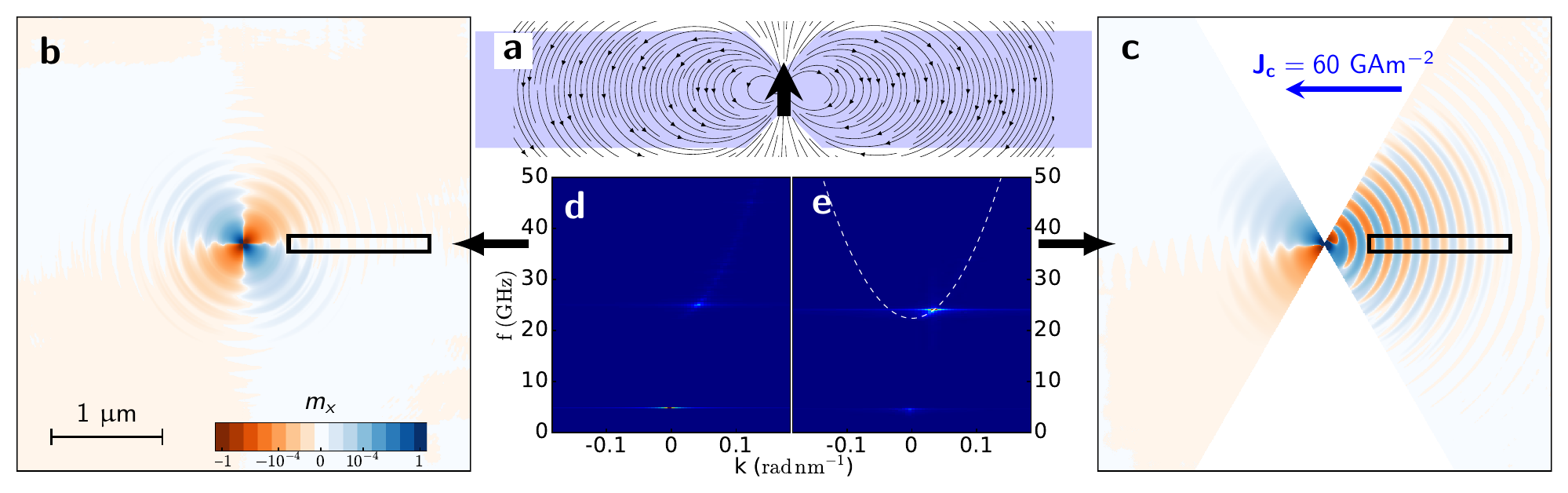}

\caption{\textbf{a} Schematic representation of the spatial configuration of the in-plane stray field generated by the DW. \textbf{b} Snapshot of magnetization dynamics in a thin film where excitation is produced by a dipolar field located at the film center rotating at $5 \unit{GHz}$.\textbf{d} $f-k$ diagram extracted from the region indicated in \textbf{b}, showing the non-propagating oscillation at 5 GHz and the propagating one at 25 GHz. \textbf{c} Snapshot from simulations where an applied current induces rotation of a DW pinned at the centre. \textbf{e} $f-k$ diagram shows propagation of SW.}\label{Fig:6}
\end{figure}

In order to shed more light on the excitation mechanism, we focus our attention on the role played by the stray field of the rotating DW. 
Due to the reduced width of the strip, precession of the spins in the DW takes place in a very coherent fashion, making the DW look like a dipole, as represented schematically in Fig.~\ref{Fig:6}-a, rotating in the strip plane. The stray field generated by such dipole has a strong in-plane component and it rotates at $f=f_{DW}$. 
To verify that the DW behaving as a rotating dipole can be regarded as the main mechanism that excites propagating SW, we proceed in two steps.
First we examine the SW emission induced by an external field having the spatial distribution of the magnetic field generated by a point dipole at the centre of a squared thin film of lateral dimension $4.096 \unit{\mu m}$ with same thickness and material parameters of the wire under study and uniformly magnetized out of plane.
\begin{eqnarray}
\mathbf{B}(\mathbf{r})&=&
\frac{\mu_0}{4\pi}\left(\frac{3\mathbf{r}\left(\mathbf{m}\cdot\mathbf{r}\right)}{r^5}-\frac{\mathbf{m}}{r^3}\right),\label{dipole}\\
\mathbf{m}(t)&=&m_0\cos(\omega t)\hat{\mathbf{x}}+m_0\sin(\omega t)\hat{\mathbf{y}}.
\end{eqnarray}
After a transient turbulent dynamics with incoherent emission of SW, when a stationary regime is reached, we observe an isotropic and rather weak emission of spin waves in all directions (Fig.\ref{Fig:6}-b and movie B supplementary material). When 2-dimensional Fourier transform is performed on $m_x(t)$ along a $1.25 \unit{\mu m}$ long line starting 400 nm away from the centre of the square, we see in the $f-k$ diagram a spot at 5 GHz and $k=0$ corresponding to the non-propagating oscillation induced directly by the external dipolar field. The principal branch of propagating SW is also marked, with a peak at 25 GHz, meaning that emission is stronger at a frequency 5 times larger than the driving rotation rate. This means that the rotating dipolar field is responsible for excitation of SW at odd harmonics and such excitation is weak, comparable to the one observed also in the left domain in Fig. \ref{Fig:2}-a.
In order to highlight the analogy with the case of the rotating DW, we carve a very deep symmetric notch in the squared film, to have a 20 nm channel in the middle as in the nano wires under exam. We then set an up-down magnetization configuration with the DW pinned at the channel and apply a current of  $6\unit{GA\,m^{-2}}$ (uniform for simplicity) that yields a DW rotation frequency $f_{DW}\sim 4.8 \unit{GHz}$. In Fig.~\ref{Fig:6}-c a snapshot representing $m_x$ during the stationary dynamics shows a strong SW emission on the right side, while the perturbation in the left domain is much weaker, and not capable of exciting SW, as can also be observed in movie B from supplementary material. Extracting the $f-k$ diagram from the same spatial region and over the same time span as in the rotating dipole case (Fig.~\ref{Fig:6}-e), we find a spot at $f_{DW}$ and $k=0$ while the spot on the dispersion relation branch is exactly at $24\unit{GHz}=5f_{DW}$ with no additional SW emission along the branch.

\begin{figure}[b]
\includegraphics[width=\textwidth]{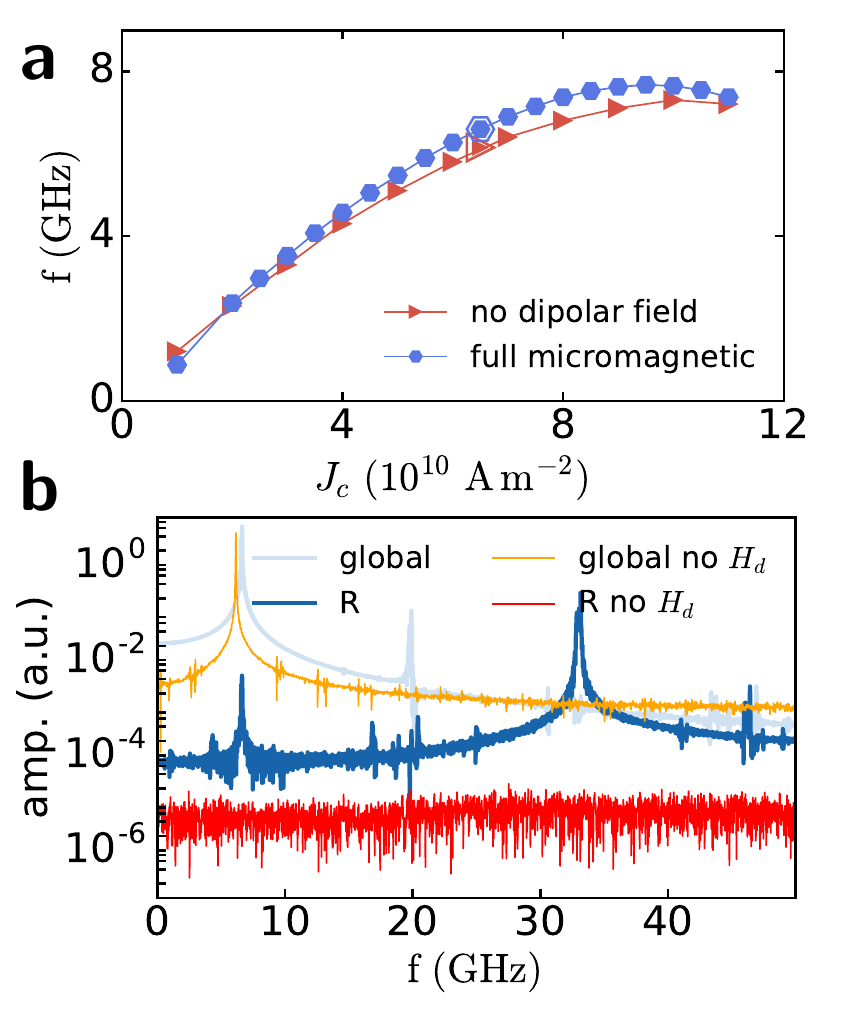}
\caption{\textbf{a} DW rotation frequency as function of current for full micromagnetic simulations (blue hexagons) compared with simulations without the non-local effect of magnetostatic field (orange triangles). 
\textbf{b} Fourier transform of $m_x(t)$ averaged over the whole sample and in the region $R$ 400 nm away from the notch as in Fig.~\ref{Fig:2}-a (light and dark blue lines) under the application of $6.5\times 10^{10}\unit{A\,m^{-2}}$. Simulations without computation of dipolar fields (orange and red lines) show a single peak at DW frequency and no signal at all away from the DW. }\label{Fig:7}
\end{figure}

\begin{figure}
\includegraphics[width=\textwidth]{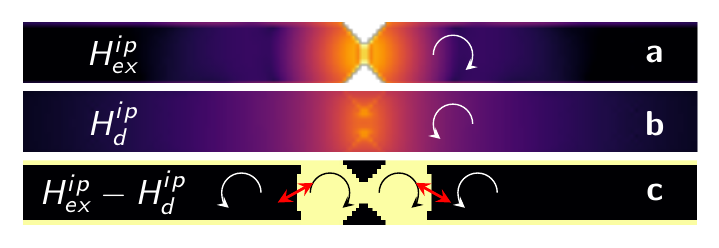}
\caption{Intensity of the in-plane component of exchange field (\textbf{a}), dipolar field (\textbf{b}) for a DW placed at the center of the wire at rest. The arrow indicates the direction of rotation of the in-plane component of the field when the DW is led to rotation via an applied current. \textbf{c} The sign of $H_{ex}^{ip}-H_d^{ip}$ marked as bright (dark) for positive (negative). Arrows indicate the direction of rotation of the combined in-plane excitation. }\label{Fig:8}
\end{figure}

If we now run micromangetic simulations without considering the long range dipolar interaction, i.e. considering an anisotropy parameter $k_\mathrm{eff}=k_u-\frac{1}{2}\mu_0 M_s^2$ including the local demagnetizing effect of dipolar field, we can achieve DW pinned rotation with a behaviour and rotation frequency very similar to the one observed in full simulations (Fig.~\ref{Fig:7}-a). However, no SW emission is observed, as shown in Fig.~\ref{Fig:7}-b where the frequency spectrum at $J_c=6.5\times 10^{10} \unit{A\,m^{-2}}$ is compared with standard simulations.
From these considerations we can conclude that the dipolar field of the DW is the responsible for exciting SW, behaving like a rotating antenna, but additional contributions due to wire edges need to be considered in order to explain unidirectionality.
When neglecting dipolar interaction, the DW rotation excites circular oscillations in the spins close to the DW only via exchange interaction passing from Bloch to N\'{e}el configuration at $f_{DW}$ frequency. Such perturbation is very strong close to the DW and decays exponentially with distance from it. The DW dipolar field, on the other hand, has a magnitude that decays as $|\mathbf{r}|^{-3}$ from the DW. In Fig.~\ref{Fig:8}-a and b the magnitude $H=\sqrt{H_x^2+H_y^2}$ of the in-plane component of the two fields is shown in dark to bright color scale for the equilibrium configuration before applying the external current, when the DW is pinned at the centre of the notch pointing upward and no propagating SW perturb the configuration.
Moreover, when the DW is set to rotation by the application of current, the in-plane component of the two fields rotates in-plane in opposite directions: clockwise $H_{ex}^\mathrm{ip}$ following DW rotation and anticlockwise $H_d^\mathrm{ip}$, so that there is a competition between the excitation on the spins due to exchange interaction close to the DW and that due to dipolar interaction further away from it. 
Their combined effect results in a strongly elliptical excitation of magnetization in the region where the two fields have similar magnitude, as shown in Fig. ~\ref{Fig:8}-c. 

\begin{figure}
\includegraphics[width=\textwidth]{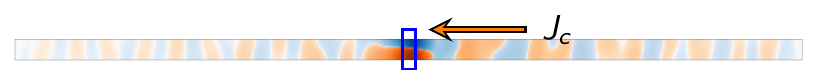}
\caption{Snapshot of magnetization dynamics showing $x-$  component of magnetization when a DW is forced to rotate via the application of a $9\times10^{10}\unit{A\,m^{-2}}$ current density. The pinning is realized via a decrease of $20\%$ in uniaxial anisotropy constant $k_u$ in the framed region.}\label{Fig:9}
\end{figure}

At the wire edges, where in-plane tilt of the spins produces surface charges and additional stray field component, this effect is strengthened. The fact that the DW is pushed by STT from the center of the strip towards the right makes the excitations at the edges much weaker on the left side where both fields have small in-plane components so that their interaction is not capable to excite higher harmonics. To prove this point, simulations with a different DW pinning strategy have been performed. 
A $20\%$ lower uniaxial anisotropy constant $k_u$ in a 30 nm wide band at the centre of the nanowire (indicated by the blue rectangle in Fig.~\ref{Fig:9}) creates an energetically favourable position for the DW, giving rise to a strong and localized potential well for the DW without changing the local geometry. Applying a current that can produce pinned rotation as shown in Fig.~\ref{Fig:9} produces SW emission in both directions. This also proves that the small non-adiabatic torque we use does not play a role in suppressing spin waves that propagate against electron flow~\cite{Seo2009}.
From this we can conclude that the dipolar field of the rotating DW is the responsible for the higher harmonic SW emission and since this excitation is of dipolar and thus geometrical origin, the displacement of the DW on the right side of the pinning site causes the screening of the emission towards the left side. Such emission is recovered when $f_{DW}>f_0$ and the simple DW rotation can excite SW towards the left (see Fig.~\ref{Fig:2}-a and \ref{Fig:6}-d).

In conclusion, we have presented a novel paradigm to excite spin waves via the STT induced rotation of a domain wall pinned at a geometrical constriction in a narrow wire. We have shown that by selecting notch shape and wire cross section the operating window of the device can be optimized and that inside the operating window spin wave emission in the direction of electrons' flow is observed at an odd multiple of DW rotation frequency up to 40 GHz without any external applied field. Such spin wave emission is very coherent and of short wavelength ($\lambda<120$ nm) typical of exchange spin waves. 
The application of an external field opposing STT has the twofold effect of extending the operation window of the DW pinned rotation regime achieving higher DW rotation frequencies on one hand, and affecting anti-symmetrically the dispersion relation in the two domains on the other, modulating the SW emission in the direction of electrons flow and allowing propagation in the opposite direction. This means that such SW emitter can work as unidirectional or asymmetric bidirectional SW emitter depending on the application of an adequate external field. 
The dipolar field of the rotating DW is the main cause of periodic non-linear excitation of SWs that propagate at higher harmonics of DW rotation frequency in the system. The displacement of the DW on one side of the notch enhances the excitation on one side and weakens it on the other, giving rise to the unidirectionality.

\section*{Methods}

\subsection*{Micromagnetic simulations}
In our study, we integrate numerically, using a custom finite difference solver, the Landau-Lifschitz-Gilbert (LLG) equation of magnetization dynamics that includes the contribution of spin transfer torque due to the flowing of an in-plane charge current density $\mathbf{J}_c$~\cite{Zhang2004} with spin polarization $P$ and degree of non-adiabaticity $\beta$
\begin{eqnarray}
\dmdt&=&-\gamma_0\m\times\heff+\alpha\m\times\dmdt-\left(\mathbf{u}\cdot\nabla\right)\m+
\beta\m\times\left[\left(\mathbf{u}\cdot\nabla\right)\m\right],\label{LLG00}
\end{eqnarray}
Here $e$ is the negative electron charge, $\mu_B$ is Bohr magneton, $\gamma_0=2.21\times10^5\unit{rad\,m\,A^{-1}s^{-1}}$  is the gyromagnetic ratio, $\alpha$ is Gilbert's damping parameter and $\mathbf{u}\stackrel{\mathrm{def}}{=}\mathbf{J}_c\frac{P\mu_B}{e M_s}$.

Material parameters of annealed 1 nm thick
$\mathrm{Co_{20}Fe_{60}B_{20}}$
as in \cite{HerreraDiez2015} have been chosen: 
saturation magnetization $M_s=8.84\times 10^5\unit{A m^{-1}}$, uniaxial anisotropy constant $k_u=8.35\times 10^5\unit{J m^{-3}}$, exchange stiffness $A_\mathrm{ex}=23\times 10^{-12}\unit{J m^{-1}}$,  Gilbert's damping $\alpha=0.015$.

The degree of non-adiabaticity of spin transfer torque has been chosen as $\beta=2\alpha$ and polarization coefficient $P=0.5$. The CoFeB strip under study is divided in micromagnetic squared cells of 4 nm in side and 1 nm thick, all dimensions below the Bloch length $\sqrt{A/k_u}=5.25\unit{nm}$.  
Absorbing boundary conditions are applied at the wire ends in the form of a smoothly augmented damping profile~\cite{Consolo2007}, in order to avoid reflection of SW and simulate propagation in a much longer nanowire.  The spatial configuration of current density $\mathbf{J}_c$ is computed numerically as the divergence of the scalar potential diffused between two contacts at the edges of the wire integrating Laplace's equation. LLG equation was integrated using a Runge-Kutta, Dormand-Prince predictor-corrector algorithm~\cite{Dormand1986} with embedded error control.
We initialize a pinned Bloch DW starting configuration. Simulation are run for 15 ns without saving output to skip the initial turbulent dynamic. Afterwards, simulations run for 40 ns. Output is written every 5 ps.

\subsection*{One dimensional model for pinned DW rotation}

Domain wall dynamics in nano-wires is well described by the so called one dimensional analytical model\cite{Schryer1974,Mougin2007a} with the inclusion of STT\cite{Zhang2004,Thiaville2004,Beach2008}. In its simplest form the model takes into account the DW position $q$ and the in-plane orientation $\phi$ of the DW spins. We make use of this model to derive the conditions that need to be fulfilled to achieve DW pinned rotation in our system.
The two differential equations describing the dynamics of a DW moving along a wire with a geometrical constriction are, in explicit form~\cite{Bisig2009,Martinez2011}
\begin{eqnarray}
\dot{q}&=&\frac{\gamma_0\Delta}{1+\alpha^2}\left( \alpha (H_p(q)+H_a) +\frac{H_K}{2}\sin2\phi \right)+\frac{1+\alpha\beta}{1+\alpha^2}u,\label{dotQPin0}\\
\dot{\phi}&=&\frac{\gamma_0}{1+\alpha^2}\left(  H_p(q)+H_a-\alpha\frac{H_K}{2}\sin2\phi \right)+\frac{\beta-\alpha}{1+\alpha^2}\frac{u}{\Delta}.\label{dotPPin0}
\end{eqnarray}
Where $H_K=2K_\perp/\mu_0 M_s$ is the in-plane shape anisotropy field on the DW, $\Delta$ is the DW width parameter, $u=|\mathbf{u}|$, $H_a$ is an external field applied along $\mathbf{\hat{z}}$
 and $H_p(q)=-{(2\mu_0 M_s L_yL_z)}^{-1}\frac{\partial V_p(q)}{\partial q}$
represents the pinning field due to a geometrical constriction at position $q=0$. 
We approximate the effect of the geometrical constriction as a parabolic potential well centred at the notch ($q=0$) with stiffness $k \unit{(J\,m^{-2})}$ and width $\ell$
\begin{equation*}
V_p(q)=\begin{cases}
		\frac{k}{2}q^2  & \text{ if } |q|<\ell\\
		0 & \text{ else}
		\end{cases}.
\end{equation*}

\section*{Acknowledgements}
This work was was supported by project WALL (FP7- PEOPLE-2013-ITN 608031) from the European Commission, project MAT2014-52477-C5-4-P from Spanish government and projects SA282U14 and SA090U16 from Junta de Castilla y Leon.
%

%
%
%
%
%
%

%

\clearpage

\begin{figure}[b]
\includegraphics[width=\textwidth]{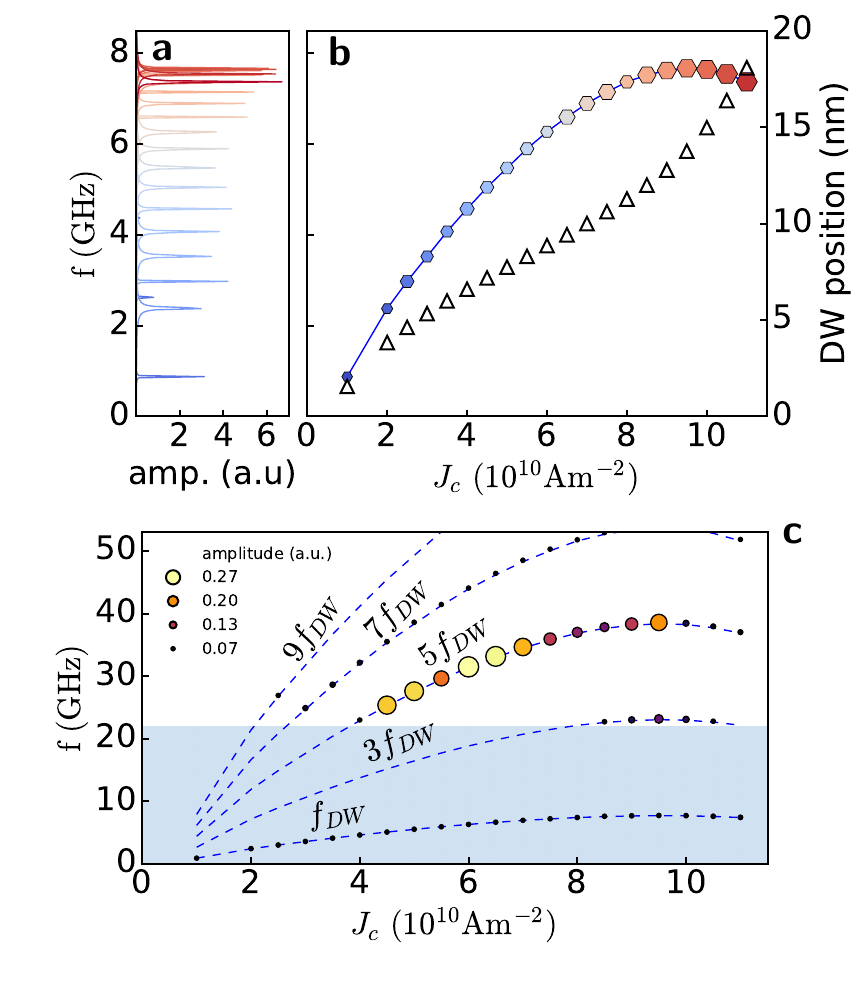}
\caption{\textbf{a} Frequency spectra of $\langle m_x \rangle$ showing the peaks from which $f_{DW}$ is extracted. \textbf{b} DW frequency (full hexagons) extracted from the peaks in \textbf{a} as function of applied current. Average DW position in the pinned rotation regime for the corresponding current (triangles). \textbf{c} Principal peaks in the frequency spectrum extracted from region $R$ away from the DW. Peak amplitude is denoted by circle size and color scale, (dark to bright). Dashed lines denote $f_{DW}$ as in \textbf{b} and its odd multiples. Shaded region denotes the non-propagating frequency gap $f<f_0$.
}\label{Fig:4}
\end{figure}

\clearpage

\section*{Supplementary Material}

\subsection{One dimensional model for pinned DW rotation}
In this section we use the one dimensional model~\cite{Schryer1974,Mougin2007a,Thiaville2004} to derive analytical expressions for the critical currents $J_\mathrm{rot},J_\mathrm{dep}$, bounds of the operating window of a pinned DW oscillator.

We start from the one dimensional equations of dynamics in the absence of an external applied field,
\begin{eqnarray}
\dot{q}&=&\frac{\gamma_0\Delta}{1+\alpha^2}\left( \alpha H_p(q) +\frac{H_K}{2}\sin2\phi \right)+\frac{1+\alpha\beta}{1+\alpha^2}u\label{dotQPin}\\
\dot{\phi}&=&\frac{\gamma_0}{1+\alpha^2}\left(  H_p(q)-\alpha\frac{H_K}{2}\sin2\phi \right)+\frac{\beta-\alpha}{1+\alpha^2}\frac{u}{\Delta}.\label{dotPPin}
\end{eqnarray}
The DW will not propagate as long as the pinning restoring force compensates the drive of STT in equation~\eqref{dotQPin}. This means that as long as
\[
u =u_\mathrm{dep}< \frac{\gamma_0\Delta}{2(1+\alpha\beta)}\left(\frac{\alpha k \ell}{\mu_0 M_s L_y L_z}\right)
\]
there exists a position $\overline{q}$ such that $\gamma_0\Delta\alpha H_p(\overline{q})=-(1+\alpha\beta)u$. 

Substituting $H_p(\overline{q})$ with this condition in \eqref{dotPPin} we obtain 
\begin{equation}
\dot{\phi}=-\frac{u}{\alpha\Delta}-\frac{\alpha\gamma_0}{(1+\alpha^2)}\frac{H_K}{2}\sin 2\phi.\label{frequency}
\end{equation}

Equilibrium in the system, with no rotation of the DW $\dot{\phi}=0$, is reached if $\sin 2\phi^*=-\frac{2u}{\gamma_0\Delta H_K}$ which means 
\[\phi^*=\frac{1}{2}\arcsin\left(-\frac{2u}{\gamma_0\Delta H_K}\right)\]
only possible if $|u|\leq |\frac{\gamma_0\Delta H_K}{2}|=u_\mathrm{rot}$.

In the end, we have two bounding conditions for the working window of the DW oscillator $u_\mathrm{rot}< u < u_\mathrm{dep}$ provided that $u_\mathrm{rot}<u_\mathrm{dep}$. The parameters playing a role in the extent of the working window are the shape anisotropy $H_K$ which has to be minimized to minimise $u_\mathrm{rot}$ and the pinning strength $k$ and extent $\ell$ which have to be maximized to have a large $u_\mathrm{dep}$.

On the other hand, if we consider the effect of a uniform external applied field as in the last part of the results section, we can extract the condition for which pinning and external field both along $-z$ equilibrate the STT and, thus, keep the DW pinned for higher applied currents. From
\begin{eqnarray}
\dot{q}&=&\frac{\gamma_0\Delta}{1+\alpha^2}\left( \alpha \left(H_a+H_p(q)\right) +\frac{H_K}{2}\sin2\phi \right)+\frac{1+\alpha\beta}{1+\alpha^2}u=0
\end{eqnarray}
 we obtain
\begin{equation}
u_D=-\frac{\alpha\gamma_0\Delta}{1+\alpha\beta}\left(H_p(\overline{q})+H_a\right).
\end{equation}

Using $u=\frac{J_c P \mu_B}{e M_s}$ we have
\begin{equation}
J_\mathrm{dep}(H_a)=\frac{e M_s}{P \mu_B}\frac{\alpha\gamma_0\Delta}{1+\alpha\beta}\left(H_p(\overline{q})+H_a\right)=J_\mathrm{dep}^0+\frac{e M_s}{P \mu_B}\frac{\alpha\gamma_0\Delta}{1+\alpha\beta}H_a,
\end{equation}

where $J_\mathrm{dep}^0$ is the threshold depinning current at zero applied field.

\subsection{Non-uniform current density effects on DW dynamics}
\begin{figure}[!h]
\includegraphics[width=0.8\textwidth]{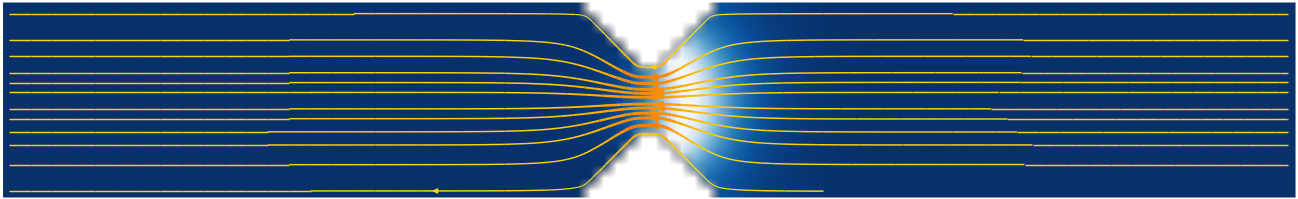}
\caption{Streamplot representing the current density flux at the constriction, the intensity is represented in color scale. A typical DW position during pinned rotation is shown in bright color in the background. }\label{current_distribution}
\end{figure}

The current density is expected to spatially vary due to the presence of a constriction along the wire. Its spatial configuration is computed numerically and represented in Fig.~\ref{current_distribution} as stream lines.

\begin{figure}[!h]
\includegraphics[width=0.8\textwidth]{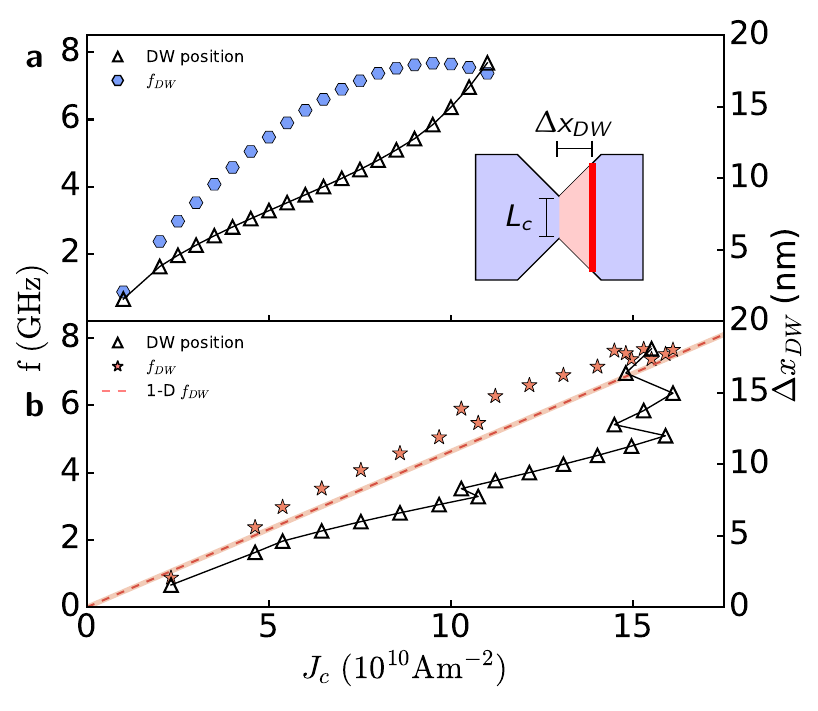}
\caption{\textbf{a} DW rotation frequency and position as function of nominal current density. Inset: schematic showing how $\Delta x_{DW}$ is evaluated. \textbf{b} DW rotation frequency and position as function of the current density measured at the centre of the DW. Dashed line is the analytical prediction from equation~\eqref{frequency}.}\label{fVSjReal}
\end{figure}

The spatial variation of current density introduces additional complexity to the problem. In fact, local current density at the DW is maximum when it is located at the centre of the notch and it decreases as the DW is pushed away from the centre,  as shown in Fig.~\ref{current_distribution}. This is clarified by looking at DW rotation frequency and average position as function of the current density measured at the DW position as it is done in Fig.~\ref{fVSjReal}-b. As the current density increases, the DW moves further away from the notch where current density is lower. 
As can be observed, if we take the data from Fig.~\ref{fVSjReal}-a (same as Fig.\ref{Fig:4}-b) and plot them against the current $J_c$ flowing at the DW position, we obtain at first a linear increase of $f_{DW}$ with $J_c$ as predicted by the analytical model~\eqref{frequency}.
Above 8.5 $\times10^{10} \unit{A\,m^{-2}}$ however, the further displacement of the DW from the centre of the notch and concurrent reduction of the local current density yields a stabilization of the effective current density flowing at the DW, yielding an almost constant $f_{DW}$. Average
DW position is computed from micromagnetic simulations as 
$
\Delta x_{DW}=-\frac{L_c}{2}+\sqrt{\frac{L_c^2}{4}+A_{tot}\langle m_z \rangle}
$
representing the height of the trapezoidal region that reversed magnetization assuming the DW as a straight line, as schematically shown in inset in Fig.~\ref{fVSjReal}-a, and $A_{tot}$ is the total upper surface of our sample.

\end{document}